\newcommand{\be}{\begin{equation}}
\newcommand{\ee}{\end{equation}}
\newcommand{\bea}{\begin{eqnarray}}
\newcommand{\eea}{\end{eqnarray}}

\def\vass#1{\left\vert\ #1 \right\vert}

\def\derp#1#2{\rp{\partial{#1}}{\partial{#2}}}
\def\rfr#1{eq.(\ref{#1})}

\def\Rfr#1{Eq.(\ref{#1})}

\def\dert#1#2{\frac{d#1}{d#2}}
\def\eqi{\begin{equation}}
\def\eqf{\end{equation}}
\def\eqia{\begin{eqnarray}}
\def\eqfa{\end{eqnarray}}

\def\rp#1#2{{#1\over#2}}

\def\ct#1{\cite{#1}}
\def\lb#1{\label{#1}}

\documentclass[11pt]{article}

\usepackage{amsmath,amsthm,amscd,amssymb}
\usepackage{latexsym}
\usepackage{graphics,graphicx}

\begin{document}

\noindent{\bf \LARGE{On the Possibility of Measuring the
Gravitomagnetic Clock Effect in an Earth Space-Based Experiment }}
\\
\\
\\
 Lorenzo Iorio\\Dipartimento Interateneo di Fisica dell'
Universit${\rm \grave{a}}$ di Bari
\\Via Amendola 173, 70126\\Bari, Italy
\\
\\
Herbert I.M. Lichtenegger\\Institut f${\rm \ddot{u}}$r
Weltraumforschung,\\ ${\rm {\ddot{O}}}$sterreichische Akademie der
Wissenschaften,\\ A-8042 Graz, Austria

\begin{abstract}
In this paper the effect of the post-Newtonian gravitomagnetic
force on the mean longitudes $l$ of a pair of counter-rotating
Earth artificial satellites following almost identical circular
equatorial orbits is investigated and the possibility of measuring
it is examined. The observable is the difference of the times
required for $l$ to pass from 0 to 2$\pi$ for both senses of
motion. Such a gravitomagnetic time shift, which is independent of
the orbital parameters of the satellites, amounts to 5$\times
10^{-7}$ s for the Earth; it is cumulative and should be
measurable after a sufficiently high number of revolutions. The
major limiting factors are the unavoidable imperfect cancellation
of the Keplerian periods, which yields a constraint of 10$^{-2}$
cm in knowing the difference between the semimajor axes $a$ of the
satellites, and the difference $I$ of the inclinations $i$ of the
orbital planes which, for $i\sim 0.01^\circ$, should be less than
$0.006^\circ$. A pair of spacecraft endowed with a sophisticated
intersatellite tracking apparatus and drag-free control down to
10$^{-9}$ cm s$^{-2}$ Hz$^{-\frac{1}{2}}$ level might allow to
meet the stringent requirements posed by such a mission.
\end{abstract}

\newpage
\tableofcontents
\newpage
\section{Introduction}
\begin{figure}
\begin{center}
\includegraphics{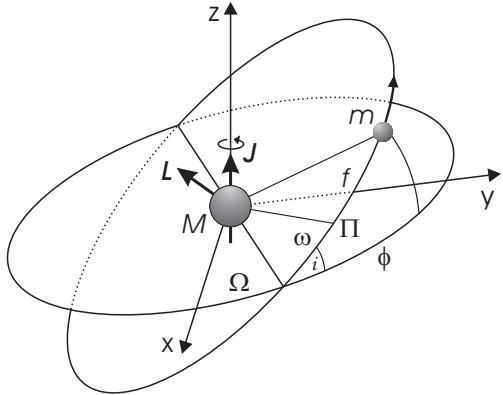}
\end{center}
\caption{\label{figura1}Orbital geometry for a motion around a
central mass. Here $L$ denotes the orbital angular momentum of the
particle of mass $m$, $J$ is the proper angular momentum of the
central mass $M$, $\Pi$ denotes the pericentre position, $f$ is
the true anomaly of $m$, which is counted from $\Pi$, $\Omega$,
$\omega$ and $i$ are the longitude of the ascending node, the
argument of pericentre and the inclination of the orbit with
respect to the inertial frame $\{x,y,z\}$ and the azimuthal angle
$\phi$ is the right ascension counted from the $x$ axis.}
\end{figure}

The orbital path of a test particle freely falling in the
gravitational field of a central body of mass $M$ and angular
momentum $J$ is affected by various post-Newtonian general
relativistic effects  of order $\mathcal{O}(c^{-2})$\ct{ciuwhe95,
mashspanish} like the Einstein gravitoelectric secular precession
of the argument of pericentre\footnote{For an explanation of the
Keplerian orbital elements of the orbit of a test particle see,
e.g., \ct{kau66} and Figure \ref{figura1}.} $\omega$ \ct{EIN} and
the Lense-Thirring gravitomagnetic secular precessions of the
longitude of the ascending node $\Omega$ and the argument of the
pericentre $\omega$ \ct{LT}.
The Einstein precession is now known at a $10^{-4}$ level of
relative accuracy, e.g.,  from Solar System planetary motions by
analyzing more than 280,000 position observations (1913-2003) of
different types including radiometric observations of planets and
spacecrafts, CCD astrometric observations of the outer planets and
their satellites, meridian transits and photographic observations
\ct{pit03}. On the contrary, to date there are not yet clear and
undisputable direct tests of the various effects induced by the
gravitomagnetic field. In April 2004 the Gravity Probe B (GP-B)
mission \ct{eve} has been launched. Its goal is to measure, among
other things, the gravitomagnetic precessions of the spins of four
superconducting gyroscopes \ct{schiff} carried onboard with a
claimed accuracy of 1$\%$ or better. The Lense-Thirring effect on
the orbits of the LAGEOS and LAGEOS II Earth's artificial
satellites has been experimentally checked for the first time some
years ago with a claimed accuracy of the order of 20$\%$
\ct{ciuff}, but, at present, there is not yet a full consensus on
the real accuracy obtained in that test \ct{ries}.

Recently, it has been shown that also the orbital period of a test
particle is affected by the post-Newtonian gravitoelectromagnetic
forces. The gravitomagnetic correction to the period of the
(equatorial) right ascension\footnote{It is nothing but the
azimuth angle $\phi$ of a spherical coordinate system in a frame
whose origin is in the center of mass of Earth, the $\{x,y\}$
plane coincides with the Earth equatorial plane and the $x$ axis
points toward the Vernal Equinox $\curlyvee$ (see Figure
\ref{figura1}).  In satellite dynamics it is one of the direct
observable quantities \ct{beutetal03}. } $\alpha$, for $e=0$,
$i=0^\circ$, where $e$ and $i$ are the eccentricity and the
inclination, respectively, of the test particle's orbit, has been
worked out in \ct{gvm}. The case of more general orbits has been
treated in \ct{iorclock} ($e=0$, $i\neq 0^\circ$) and
\ct{mashclock} ($e\neq 0$, $i\neq 0^\circ$). The gravitoelectric
correction has been treated in \ct{mashclock}. Let us consider a
pair of counter-orbiting satellites, conventionally denoted as
$(+)$ and $(-)$, following identical circular and equatorial
orbits along opposite directions around a central spinning body of
mass $M$ and proper angular momentum $J$; while the
gravitoelectric correction to the Keplerian period has the same
sign for both the directions of motion, similar to the classical
perturbing corrections of gravitational origin (oblateness of the
central mass, tides, N-body interactions), the gravitomagnetic one
changes its sign if the motion is reversed. This allows, at least
in principle, to single out the latter one by measuring the
difference of the periods of the counter-orbiting satellites. Some
preliminary error analyses can be found in \ct{errclock}.

As we will show here, it turns out that also the mean
longitudes\footnote{$\mathcal{M}$ is the mean anomaly of the test
particle's orbit. $l$, along with the Keplerian orbital elements,
comes out from the machinery of the data reduction process
performed by the orbit determination softwares like GEODYN II and
UTOPIA.} $l=\mathcal{M}+\Omega+\omega$ are affected, among other
perturbations, by the post-Newtonian general relativistic
gravitoelectromagnetic forces. Also in this case the
gravitomagnetic correction is sensitive to the direction of
motion, contrary to the gravitoelectric one.

In this paper we will investigate the possibility of singling out
the gravitomagnetic effect on the mean longitudes by means of a
suitable space-based mission in the Earth space environment. We
will mainly deal with a number of competing classical effects
which could represent a source of systematic errors and bias. No
measurements error budget is carried out.

\section{The Gravitomagnetic Effect on the Mean Longitude}

From the Gauss perturbative equations \ct{milnobfar87} it turns
out that the rate equation for $l$, for small but finite values of
$e$ and $i$, is \eqi\dert{l}{t}\sim
n-\frac{2}{na}A_R\left(\frac{r}{a}\right)+\left(\frac{i^2}{2}\right)\dert{\Omega}{t}+\left(\frac{e^2}{2}\right)
\dert{\omega}{t},\label{piccoli}\eqf where $a$ is the satellite's
semimajor axis and $n=(GMa^{-3})^{\frac{1}{2}}$ is the Keplerian
mean motion; $A_R$ is the radial component of the perturbing
acceleration ${\bf a}_{\rm pert}$. The currently available
technologies allows to insert Earth artificial satellites in
orbits with\footnote{This value holds for initial orbital
injection. Later, various perturbations, mainly due to the
terrestrial gravitational field, affect $e$ which can reach values
up to $5\times 10^{-3}$. } $e\lesssim 10^{-3}$ and $i\sim
0.01^\circ$; then, the third and fourth terms in \rfr{piccoli},
which are proportional to $i$ and $e$ because $d\Omega/dt\propto
1/\sin i$ and $d\omega/dt\propto 1/e$, can be neglected and the
equation \eqi\dert{l}{t}\sim n-\frac{2}{na}A_R\label{longi}\eqf
can be used instead of \rfr{piccoli} to a good level of
approximation.

According to \ct{joograf91}, the radial components of the
post-Newtonian general relativistic gravitoelectromagnetic
accelerations, for generic orbits around a central spinning body,
are
\begin{eqnarray}
A_R^{\rm (GE)}&=& \frac{(GM)^2(1+e\cos f)^2}{c^2 a^3 (1-e^2)^3}(3+2e\cos f -e^2 + 4e^2\sin^2 f),\label{rge}\\
A_R^{\rm (GM)}&=&\rp{2nGJ\cos i}{c^2
a^2(1-e^2)^{\rp{7}{2}}}(1+e\cos f)^4 ,\lb{rlt}
\end{eqnarray}
where $f$ is the true anomaly. They are induced by the
post-Newtonian general relativistic gravitoelectromagnetic
fields\footnote{In the weak-field and slow-motion approximation of
the General Theory of Relativity the equations of motion of a test
particle are \ct{mashspanish} \eqi\frac{d^2{\bf r}}{dt^2}=-{\bf
E}_{\rm g}-{\bf E}^{\rm (GE)}-2\frac{{\bf v}}{c}\times{\bf B}^{\rm
(GM)},\eqf where ${\bf E}_{\rm g}=GM{\bf r}/r^3$ is the usual
Newtonian monopole term. }
\begin{eqnarray}
{\bf E}^{\rm (GE)}&=&-\frac{GM}{c^2r^3}\left(\frac{4GM}{r}-{\bf
v}^2\right)\bold{r}-\frac{4GM}{c^2r^2}(\bold{r}\cdot{\bf v}){\bf
v},\label{grem}\\
{\bf B}^{\rm (GM)}&=&-\frac{GJ}{cr^3}[{\bf \hat{J}}-3({\bf \hat
J}\cdot{\bf \hat r}){\bf\hat{r}}]. \lb{bgm}
\end{eqnarray}

From \rfr{rge}-\rfr{rlt} it can be noted that, while the
gravitoelectric acceleration is insensitive to the sense of motion
of the test particle along its orbit, it is not so for the
gravitomagnetic acceleration due to its dependence on $\cos i$.
Indeed, according to \ct{vinti}
\begin{eqnarray} \sin i & = & \rp{(h_x^2+h_y^2)^{\rp{1}{2}}}{\vass{\bf h}},\\
 \cos i & = & \rp{h_z}{\vass{\bf h}},\end{eqnarray}
 where ${\bf h}$ is the
 orbital angular momentum per unit mass whose components change sign when ${\bf v}\rightarrow -{\bf v}$.
 By reversing the sign of the velocity vector one obtains ($+$ and $-$ denote the pro- and retrograde orbits, respectively)
\begin{eqnarray}\sin i^{(-)} & = & \sin i^{(+)},\\
\cos i^{(-)} & = & -\cos i^{(+)}, \lb{inkz}
\end{eqnarray}
  from which it follows
  \eqi i^{(-)} = 180^\circ-i^{(+)}.\eqf

By inserting \rfr{rge}-\rfr{rlt} in \rfr{longi}, it can be
obtained for the time $P_l$ required to $l$ for passing from 0 to
2$\pi$
\begin{eqnarray}
P_l^{(\pm)}:=P_l^{(0)}+P_l^{(\rm GE)}+P_l^{(\rm
GM)}=\frac{2\pi}{n}+\frac{12\pi(GM
r_0)^{\frac{1}{2}}}{c^2}\pm\frac{8\pi J}{c^2 M}.\label{plong}
\end{eqnarray}
\Rfr{plong} yields for identical orbits followed in opposite
directions, \eqi\Delta P_l:=P_l^{(+)}-P_l^{(-)}=\rp{16\pi J}{c^2
M},\label{DPGM}\eqf which amounts to $5\times 10^{-7}$ s for Earth
and is four times larger than the corresponding effect for the
right ascension.

\section{Sources of Systematic Errors}

\Rfr{DPGM} would be valid only if the Keplerian periods (and the
various classical and post-Newtonian gravitoelectric perturbative
corrections to them) of the two satellites were exactly equal;
this condition, {\rm however}, cannot be achieved due to the
unavoidable orbital injection errors. Then, \rfr{DPGM} has also to
account for the difference induced in the Keplerian periods and
the various classical and post-Newtonian gravitoelectric
corrections to them, e.g., by the difference $d$ in the semimajor
axes of the two satellites. The differences in $i$, which do not
affect the Keplerian periods and the post-Newtonian
gravitoelectric correction, would have, instead, an impact on the
classical perturbative terms which cannot be neglected.

In general, the difference between the mean longitude periods of
the two counter-rotating satellites can be written as \eqi \Delta
P_l=\Delta P^{(0)}+\Delta P_l^{\rm class}+\Delta P_l^{(\rm GE)}
+\Delta P_l^{(\rm GM)}.\lb{herli}\eqf Over many orbital
revolutions, say $N$, the accuracy in determining the
gravitomagnetic time shift becomes \eqi\delta[\Delta P^{(\rm
GM)}]\leq\rp{\delta (\Delta P)^{\rm exp}}{N}+\delta[\Delta
P^{(0)}]+\delta[\Delta P_l^{\rm class}]+\delta[\Delta P_l^{\rm GE
}],\eqf where $\delta (\Delta P_l)^{\rm exp} $ is the experimental
error in the difference of the obtained $P_l^{(\pm)}$ over $N$
revolutions. After a sufficiently high number of orbital
revolutions it should be possible to make the term ${\delta
(\Delta P_l)^{\rm exp}}/{N}$ smaller than $\Delta P_l^{(\rm GM)}$.
In order to get an estimate, let us calculate the angular shift
corresponding to $\Delta P_l^{(\rm GM)}$ over an orbital
revolution by using \eqi\frac{\Delta
l}{\overline{l}}\sim\frac{\Delta P_l^{(\rm GM)}}{P_l^{(0)}}.\eqf
For\footnote{In order just to fix the ideas, we shall consider the
semimajor axis of the existing SLR ETALON satellites. We will see
that high altitudes contribute to reduce the impact of many
aliasing perturbations. } $r_0=25498$ km and
$P_l^{(0)}=4.05200895378\times 10^4$ s, $\overline{l}=2\pi$ the
gravitomagnetic shift amounts to $\Delta l=1\times 10^{-2}$
milliarcseconds (mas in the following; 1 mas = 4.8$\times 10^{-9}$
rad), over an angular span. Since the accuracy with which the
orientation of the axes of the International Terrestrial
Reference\footnote{Its realization, the International Terrestrial
Reference Frame (ITRF), is obtained by estimates of the
coordinates and velocities of a set of observing stations on the
Earth \ct{mc}.} System (ITRS) is of the order\footnote{See on the
Internet http://hpiers.obspm.fr/iers/bul/bulb/explanatory.html} of
3 mas, after 300 orbital revolutions or 140 days, the
gravitomagnetic time shift would reach this sensitivity cutoff.

\subsection{The Impact of the Imperfect Cancellation of the Keplerian Periods}

As we will see, the major limiting factor in measuring the
gravitomagnetic time shift of interest is the difference of the
Keplerian orbital periods \eqi \Delta
P^{(0)}=\rp{2\pi[r_0^{\rp{3}{2}}-(r_0+d)^{\rp{3}{2}}]}{(GM)^{\rp{1}{2}}}\sim
\rp{3\pi d r_0^{\rp{1}{2}}}{(GM)^{\rp{1}{2}}}
,\lb{kepldiff}\eqf where $r_0$ represents the nominal value of the
semimajor axis of the two satellites. The shift of \rfr{kepldiff}
cannot be made smaller than \rfr{DPGM}, in practice, by choosing
suitably the orbital geometry of the satellites. Indeed, from
\rfr{kepldiff} follows that the relation
$d{r_0}^{\rp{1}{2}}\leq 1.059\times 10^3$ cm$^{\frac{3}{2}}$
should be fulfilled; for $r_0\sim 10^9$ cm it would imply $d\sim
3\times 10^{-2}$ cm. Then, in order to be able to measure the
relativistic effect of interest, which accumulates {\rm during
the} orbital revolution, the difference  should be subtracted from
the data provided that its error $\delta[\Delta P^{(0)}]$, which
is present at every orbital revolution and is due to the
uncertainties in the Earth $GM$, $r_0$ and $d$, is smaller than
the gravitomagnetic time shift. {\rm This error is given by}
\eqi\delta[\Delta P^{(0)}]\leq\left|\rp{3\pi dr_0^{\rp{1}{2}}
}{2(GM)^{\rp{3}{2}}}\right|\times\delta(GM)+\left|\rp{3\pi
d}{2(GMr_0)^{\rp{1}{2}}}\right|\times(\delta r_0)+\left|\rp{3\pi
r_0^{\rp{1}{2}}}{(GM)^{\rp{1}{2}}}\right|\times(\delta
d)\lb{errtk}\eqf {\rm and} by assuming $r_0=25498$ km, $(\delta
{r_0})^{\rm exp}=1$ cm and $\delta(GM)=8\times 10^{11}$ cm$^3$
s$^{-2}$ \ct{mc}, \rfr{errtk} yields \eqi \delta[\Delta
P^{(0)}]\leq(2.8\times 10^{-14}\ {\rm cm^{-1}\ s})\times
d+(2.3\times 10^{-5}\ {\rm cm^{-1}\ s})\times (\delta
d).\lb{numerk}\eqf \Rfr{numerk} tells us {\rm that} the error due
to the uncertainty in\footnote{Note that the uncertainty in the
Earth's $GM$ is a not a totally negligible limiting factor.
Indeed, for an intersatellite separation of a few kilometers the
induced bias is of the order of $10^{-8}$ s. If only one satellite
was considered the systematic error in the Keplerian period due to
$GM$ $\pi(a/GM)^{\rp{3}{2}}\delta(GM)$ would amount to $4\times
10^{-5}$ s, as pointed out in the preliminary evaluations of
\ct{iornongravcqg}. } $GM$ and, especially, $r_0$ is less than the
gravitomagnetic effect, while $\delta d$ should be at the level of
$2\times 10^{-2}$ cm. It seems to be impossible to meet this very
stringent requirement with the current SLR technology due to many
measurement errors (station errors, random errors in precession,
nutation and Earth rotation, observation errors). However, an
intersatellite tracking approach could yield better results. A
level of $(\delta d)^{\rm exp}$ of the order of $10^{-3}$ cm is
currently available with the K-band Ranging (KBR) intersatellite
tracking technology used for the GRACE mission \ct{kangetal03}. On
the other hand, it must be noted that it would not be easy to use
this approach for satellites following opposite directions. The
GRACE twins revolve in the same sense.

In regard to $(\delta d)^{\rm systematic}$, thinking about a pair
of completely passive, spherical LAGEOS--like satellites, it
should {\rm be pointed} out that for orbits with $e\sim i\sim 0$,
for which \rfr{DPGM} holds, many non--gravitational
perturbations\footnote{On the semimajor axis there are no
long--term gravitational perturbations.} affecting {\rm the
semimajor axis} vanish \ct{iornongravcqg}; the remaining ones
could be strongly constrained by constructing the two satellites
very carefully with regard to their geometrical and physical
properties. Some figures will be helpful. We will adopt the
physical properties of the existing LAGEOS satellites. The
atmospheric drag induces a decrease in the semimajor axis of
$\Delta a=-a^2 \varrho C_D S/m$ over one orbital revolution, where
$C_D$ is the satellite drag coefficient, $S/m$ is the satellite
area-to-mass ratio and $\varrho$ is the atmospheric density at the
satellite altitude. For $C_D\sim 4.9$, $S/m=7\times 10^{-3}$
cm$^2$ g$^{-1}$ and $\varrho=8.4\times 10^{-21}$ g cm$^{-3}$
(estimate of the atmospheric neutral density at LAGEOS altitude;
for $r_0=25498$ km it should be smaller, of course) the decrease
in the semimajor axis amounts to 4$\times 10^{-4}$ cm over one
orbital revolution. The impact of direct solar radiation pressure
on the semimajor axis of a circular orbit vanishes, also when
eclipses effects are accounted for. The nominal Poynting-Robertson
effect on the semimajor axis of a circular, equatorial orbit
($r_0=25498$ km) is of the order of $10^{-3}$ cm over one orbital
revolution. However, it is linearly proportional to the satellite
reflectivity coefficient $C_R$ for which a 0.5--1$\%$ mismodelling
can be assumed. The semimajor axis of a circular orbit is neither
affected neither the albedo over one orbital revolution. The same
holds also for the IR Earth radiation pressure and the solar
Yarkovsky-Schach effect (by neglecting the eclipses effects). The
Earth Yarkovsky-Rubincam effect would affect the semimajor axis of
an orbit with $i\sim 0^\circ$ by means of a nominal secular trend
less than $1\times 10^{-1}$ cm ($r_0=25498$ km) over one orbital
revolution\footnote{It would vanish if the satellite spin axis was
aligned with the $z$ axis of an Earth-centered inertial frame.}.
However, a 20-25$\%$ mismodelling can be assumed on it. The direct
effect of the terrestrial magnetic field on the semimajor axis of
a circular orbit vanishes over one orbital revolution.

Another possibility could be the use of more complex (and more
expensive) spacecraft endowed with a drag--free apparatus; it
would be helpful in suitably constraining $(\delta {r_0})^{\rm
systematic}$ and, especially, $(\delta d)^{\rm systematic}$.
Indeed, for orbits with semimajor axis of the order of 10$^9$ cm
the unperturbed Keplerian period is of the order of $P^{(0)}\sim
4\times 10^4$ s. If the maximum admissible error in knowing the
position of the satellites is of the order of $2\times 10^{-2}$ cm
then the maximum disturbing acceleration that would not mask the
gravitomagnetic clock effect, over one orbital revolution, is
$\delta d/[P^{(0)}]^2\sim 1\times 10^{-11}$ cm s$^{-2}$. But the
drag--free technologies currently under development for LISA
\ct{LISA} and OPTIS \ct{ditetal02} missions should allow to cancel
out the accelerations of non--gravitational origin down to
$3\times 10^{-13}$ cm s$^{-2}$ Hz$^{-{\frac{1}{2}}}$ and
10$^{-12}$ cm s$^{-2}$ Hz$^{-{\frac{1}{2}}}$, respectively. In our
case, for $r_0=25498$ km and an orbital frequency of $2.4\times
10^{-5}$ Hz, a $1\times 10^{-9}$ cm s$^{-2}$ Hz$^{-\frac{1}{2}}$
drag-free level would be sufficient.
More precisely, from the Gauss perturbative rate equation of the
semimajor axis it turns out that, for circular orbits, the effects
induced on $r_0$ by periodically time-varying non--gravitational
along-track accelerations with frequencies multiple of the orbital
frequency vanish over one orbital revolution. For constant
along-track accelerations $A_T$ the situation is different: over
one orbital revolution (corresponding to a frequency of the order
of 10$^{-5}$ Hz) their effect on $r_0$ would amount to \eqi\delta
r_0^{(\rm NG)}=4\pi\left(\frac{r_0^3}{GM}\right)A_T.\eqf In order
to make $\delta r_0^{(\rm NG)}=2\times 10^{-2}$ cm the maximum
value for $A_T$ would be $3.8\times 10^{-11}$ cm s$^{-2}$ in a
frequency range with $10^{-5}$ Hz as upper limit. It is a
difficult but not impossible limit to be obtained with the
drag-free technologies which are currently under development.

\subsection{The Impact of the Imperfect Cancellation of the Post-Newtonian Gravitoelectric Periods}


The perturbative correction to $P_l$ induced by the post-Newtonian
gravitoelectric acceleration, given in \rfr{DPGM},  amounts to
\eqi P_l^{(\rm GE)}=\frac{12\pi(GM r_0
)^{\frac{1}{2}}}{c^2}.\lb{buu}\eqf For an Earth high-orbit
satellite it is of the order of 10$^{-5}$ s. The orbital injection
errors in $r_0$ would yield \eqi\Delta P_l^{(\rm
GE)}\sim\frac{6\pi}{c^2}\left(\frac{GM}{r_0}\right)^{\frac{1}{2}}d,\eqf
which amounts to 10$^{-9}$ s for $r_0=25498$ km and $d=5$ km. This
result justifies, a posteriori, our choice of neglecting the small
corrections to \rfr{buu} induced by the small, but finite,
eccentricity of the orbit.

\subsection{The Impact of the Classical Gravitational Perturbations}

In this section the perturbations of gravitational origin on the
mean longitude $l$ and their impact on the measurement of $\Delta
P^{(\rm GM)}_l$ are investigated.

\subsubsection{The Geopotential} Since we are interested in effects
which are averaged over many orbital revolutions, only the zonal
harmonics of geopotential, which induces secular perturbations on
$\Omega,\ \omega$ and $\mathcal{M}$, will be considered. In the
following the Gaussian approach together with \rfr{longi} will be
used.

Let us write the zonal part of geopotential as \eqi \Phi^{\rm zon
}:=\Phi^{(0)}+\Delta\Phi^{\rm
zon}=\rp{GM}{r}-GM\sum^{\infty}_{\ell =2}
\frac{R^{\ell}_{\oplus}}{r^{\ell +1}}\ J_{\ell}\
P_{\ell}(\cos\theta).\eqf According to (6.98b) of \ct{bocpuc98},
the radial component of the perturbing acceleration, in spherical
coordinates, for an even zonal harmonic of degree $\ell$, is \eqi
A_R^{(J_{\ell})}=\derp{\Delta\Phi^{\rm zon }}{r}=(\ell +1
)\rp{GMR_{\oplus}^{\ell}}{r^{\ell +2}}\ J_{\ell}\
P_{\ell}(\cos\theta). \lb{rgeop} \eqf It is easy to see that the
odd zonal harmonics of geopotential do not affect the mean
longitude of a satellite in an equatorial orbit. Indeed, it is
well known that the odd-degree Legendre polynomials vanish for
$\theta=90^\circ$ \ct{roy88}.
%
The perturbing radial accelerations due to the even zonal
harmonics are
\begin{eqnarray}
A_R^{(J_2)}&=& -\rp{3GMR^2_{\oplus} J_2}{2r^4},\lb{RJ2}\\
A_R^{(J_4)}&=& \rp{15GMR^4_{\oplus} J_4}{8r^6},\\
A_R^{(J_6)}&=& -\rp{35GMR^6_{\oplus} J_6}{16r^8},\\\lb{RJ6}
A_R^{(J_8)}&=& \rp{315GMR^8_{\oplus} J_8}{128 r^{10}}.\lb{RJ8}
\end{eqnarray}
Inserting \rfr{RJ2}-\rfr{RJ6} in \rfr{longi} yields \eqi P_l^{\rm
obl}=-\rp{6\pi R^2_{\oplus} J_2}{(GMr_0)^{\frac{1}{2}}}+\rp{15\pi
R^4_{\oplus} J_4}{(GMr^5_0)^{\frac{1}{2}}}-\rp{35\pi R^6_{\oplus}
J_6}{4(GMr^9_0)^{\frac{1}{2}}}+\rp{315\pi R_{\oplus}^8J_8}{32(GM
r_0^{13} )^{\rp{1}{2}}}-...=\nonumber\eqf \eqi=-8.2341063\ {\rm
s}-1.9266\times 10^{-3}\ {\rm s}-2.34\times 10^{-5}\ {\rm
s}+5\times 10^{-7}\ {\rm s}+...\eqf for an orbit radius
$r_0=25498$ km. This shows that, for such a radius, just the first
four even zonal harmonics are of the same order of magnitude of
$P_l^{(\rm GM)}\sim 10^{-7}$ s.

Let us now evaluate the impact of the even zonal harmonics the
geopotential on the measurement of $\Delta P_l^{(\rm GM)}$ for a
pair of counter-orbiting satellites whose orbits differ by a small
amount $d\ll r_0$ in radius. With \eqi
\frac{1}{(r_0)^{\frac{n}{2}}}- \frac{1}{(r_0+d)^{\frac{n}{2}}}\sim
\frac{nd}{2(r_0)^{\frac{n+2}{2}}}, \eqf  the difference in the
perturbing terms due to the even zonal harmonics are, for $d=5$ km
and $r_0=25498$ km
\begin{eqnarray}
\Delta P_l^{(J_2)}&=& -\rp{3\pi R^2_{\oplus} J_2 d}{(GMr_0^3)^{\frac{1}{2}}}=-8.068\times 10^{-4}\ {\rm s},\lb{dJ2}\\
\Delta P_l^{(J_4)}&=& \rp{75\pi R^4_{\oplus} J_4 d}{2(GMr_0^7)^{\frac{1}{2}}}=-9\times 10^{-7}\ {\rm s},\\
\Delta P_l^{(J_6)}&=& -\rp{315\pi R^6_{\oplus} J_6 d
}{8(GMr_0^{11})^{\frac{1}{2}}}=-2\times 10^{-8}\ {\rm
s}\lb{dJ6},\\
\Delta P_l^{(J_8)}&=& \rp{4095\pi R^8_{\oplus} J_8 d
}{64(GMr_0^{15})^{\frac{1}{2}}}=6\times 10^{-10}\ {\rm s}.\lb{dJ8}
\end{eqnarray}
This means that only $\Delta P_l^{(J_2)}$ and $\Delta P_l^{(J_4)}$
are relevant. The uncertainty in the knowledge of the Earth $GM$
and of the even zonal harmonics yield
\begin{eqnarray}
\delta[\Delta P_l^{(J_2)}]_{GM}&=& \rp{3\pi R^2_{\oplus} J_2 d}{2(GMr_0)^{\frac{3}{2}}}\times \delta(GM)\sim 8\times 10^{-13}\ {\rm s},\\
\delta[\Delta P_l^{(J_2)}]_{J_2}&=& \rp{3\pi R^2_{\oplus}
d}{(GMr_0^3)^{\frac{1}{2}}}\times (\delta J_2)\sim 3\times
10^{-11}\ {\rm s},
\end{eqnarray}

\begin{eqnarray}
\delta[\Delta P_l^{(J_4)}]_{GM}&=& \rp{75\pi R^4_{\oplus} J_4
d}{4(GM)^{\frac{3}{2}}(r_0)^{\frac{7}{2}}}
\times \delta(GM)\sim 9\times 10^{-16}\ {\rm s},\\
\delta[\Delta P_l^{(J_4)}]_{J_4}&=& \rp{75\pi R^4_{\oplus}
d}{2(GMr_0^7)^{\frac{1}{2}}}\times (\delta J_4)\sim 3\times
10^{-11}\ {\rm s},
\end{eqnarray}

For $\delta(GM)$ and $\delta J_{\ell}$ the values of the IERS
convention \ct{mc} and of the EIGEN-2 Earth gravity model
\ct{eigen2}, respectively, have been used.

Another source of error is represented by the uncertainty in the
knowledge of the separation $d$ between the two orbits and of
their radius $r_0$. Their impact is, by assuming $\delta
d\sim\delta r_0\sim 1$ cm
\begin{eqnarray}
\delta[\Delta P_l^{(J_2)}]_{r_0}&=& \rp{9\pi R^2_{\oplus} J_2
d}{2(GM)^{\frac{1}{2}}(r_0)^{\frac{5}{2}}}\times (\delta r_0)
\sim 4\times 10^{-13}\ {\rm s},\\
\delta[\Delta P_l^{(J_2)}]_{d}&=& \rp{3\pi R^2_{\oplus}
J_2}{(GMr_0^3)^{\frac{1}{2}}}\times (\delta d)\sim 1\times
10^{-9}\ {\rm s},
\end{eqnarray}

\begin{eqnarray}
\delta[\Delta P_l^{(J_4)}]_{r_0}&=& \rp{525\pi R^4_{\oplus} J_4
d}{4(GM)^{\frac{1}{2}}(r_0)^{\frac{9}{2}}}
\times (\delta r_0)\sim 1\times 10^{-15}\ {\rm s},\\
\delta[\Delta P_l^{(J_4)}]_{d}&=& \rp{75\pi R^4_{\oplus}
J_4}{2(GMr_0^7)^{\frac{1}{2}}}\times (\delta d)\sim 2\times
10^{-12}\ {\rm s},
\end{eqnarray}

These results clearly show that, for the same semimajor axis of,
say, the ETALON SLR satellites and for not too stringent
requirements on the orbital injection errors for the separation
between the two orbits, the impact of the geopotential on the
measurement of $\Delta P_l^{(\rm GM)}$ is negligible.

Finally, let us consider what could be the impact of the secular
variations of the even zonal harmonics $\dot J_{\ell}$. For the
first even zonal harmonic the effective  $\dot J_2^{\rm
eff}\sim\dot J_2+0.371\dot J_4+0.079\dot J_6+0.006\dot
J_8-0.003\dot J_{10}...$, whose magnitude is of the order of
$(-2.6\pm 0.3)\times 10^{-11}$ yr$^{-1}$, can be considered. The
insertion of such value in \rfr{dJ2} yields $\Delta P_l^{(\dot
J_2^{\rm eff})}=1\times 10^{-11}$ s over one year, so that it can
be concluded that its effect can be safely neglected.

\subsubsection{The Tides}

Another source of potential systematic error is represented by the
solid Earth and ocean tides \ct{iortides}.

For a constituent of given frequency ${\rm f}$ of the solid Earth
tidal spectrum, which is the more effective in perturbing the
Earth satellites'orbits, the perturbation of degree $\ell$ and
order $m$ induced on $l$ is
\eqi\dert{l}{t}=n+A_{\ell m}(GM)^{\frac{1}{2}} R^{\ell
-1}_{\oplus} r_0^{-(\frac{2\ell +3}{2})}2(\ell
+1)\sum_{p=0}^{\ell}\sum_{q=-\infty}^{+\infty}F_{\ell m p}G_{\ell
p q }k_{\ell m}^{(0)}H_{\ell}^{m}\cos\gamma_{{\rm f}\ell m p q}
,\eqf where $A_{\ell m}=[(2\ell +1)(\ell -m)!/ (4\pi)(\ell + m )
!]^{\frac{1}{2}}$, $F_{\ell m p}(i)$ and $G_{\ell p q}(e)$ are the
inclination and eccentricity functions, respectively \ct{kau66},
$k_{\ell m}^{(0)}$ are the Love numbers, $H_{\ell}^{m}$ are the
tidal heights and $\gamma_{{\rm f}\ell m p q}$ is built up with
the satellite's orbital elements $\Omega$, $\omega$ and
$\mathcal{M}$, the lunisolar variables and the lag angle
$\delta_{{\rm f}\ell m}$ \ct{iortides}. Note that the dependence
on $F_{\ell m p}$ and $G_{\ell p q}$ is the same also for the
ocean tidal perturbations.

From an inspection of the explicit expressions of the inclination
and eccentricity functions \ct{kau66} and from the condition
$\ell-2p+q=0$, which must be fulfilled for the perturbations
averaged over an orbital revolution, it turns out that the
$\ell=3$ part of the entire tidal spectrum does not affect the
mean longitude of a satellite with $i=e=0$. With regard to the
$\ell=2$ constituents, only the long-period zonal ($m=0$) tides
induce non vanishing perturbations on $P_l$ for circular and
equatorial orbits. Among them, the most powerful is by far the
18.6-year lunar tide. Its effect on $P_l$ is given by\eqi
P_l^{(\rm 18.6-yr)}=\frac{3(5\pi )^{\frac{1}{2}}\ R_{\oplus}\
k^{(0)}_{20}(\bar{\rm f})\ H_2^0(\bar{\rm f})}{(GM
r_0)^{\frac{1}{2}}}\cos\gamma_{{\bar{\rm f}}2010}.\eqf For
$r_0=25498$ km it amounts to 3$\times 10^{-6}$ s. The difference
in the orbits of the two satellites would yield \eqi\Delta
P_l^{(\rm 18.6-yr)}=\frac{3(5\pi)^{\frac{1}{2}}\ R_{\oplus}\
k^{(0)}_{20}(\bar{\rm f})\ H_2^0(\bar{\rm f})\ d}{2(GM
r_0^3)^{\frac{1}{2}}}\cos\gamma_{{\bar{\rm f}}2010}\sim 6\times
10^{-10}\ {\rm s} \eqf for $d=5$ km and $r_0=25498$ km. As can be
seen, $\Delta P_l^{(\rm GM)}\gg \Delta P_l^{(\rm 18.6-yr)}$, so
that the impact of the direct tidal perturbations  on the
measurement of the gravitomagnetic clock effect on $l$ can be
neglected.

The cross-coupling between the static Earth oblateness and the
Earth tides \ct{chri88} would induce the following perturbations
on the mean longitude \eqi\dert{l}{t}=n-\frac{21 J_2 GM
R_{\oplus}^{\ell +1} A_{\ell m}
\sum_{p=0}^{\ell}\sum_{q=-\infty}^{+\infty}F_{\ell m p} G_{\ell p
q} [(\ell -2p)-m] k^{(0)}_{\ell m} H_{\ell}^m }{2 r_0^{\ell +5}
{\rm f}}\cos\gamma_{{\rm f}\ell m p q}.\eqf From an inspection of
the explicit expressions of the inclination functions (see
Appendix A) and from the condition $\ell-2p=0$, which holds for
even $\ell$, it turns out that the $\ell=2$ part of the indirect
tidal spectrum, which is the most powerful, has no effect on $l$
for an equatorial orbit. More precisely, $F_{201}$ does not vanish
for $i= 0^\circ$, but in this case, $\ell-2p=m=0$. The other
$\ell=2$ inclination functions vanish for equatorial orbits. The
same holds also for the $\ell=3$ constituents for which
$\ell-2p\neq 0$. It turns out that $F_{311}\neq 0$ for equatorial
orbits; in this case $\ell-2p=m=1$. The other $\ell=3$ inclination
functions vanish for $i=0^\circ$.

\subsubsection{The Impact of the Errors in the Inclinations}

Until now we have propagated the uncertainties in the orbit radius
by assuming for the inclination and the eccentricity their nominal
values $i=e=0$. Now we wish to fix $r_0$ and propagate the errors
in the inclination \ct{lichetal03}. As pointed out before, the
uncertainties in $i$ affect the classical perturbative corrections
to the Keplerian periods. The effects of the Earth oblateness are
the most prominent; thus, we will investigate them in order to
establish upper bounds in the admissible separation $I$ between
the two orbital planes.

The radial component of the perturbing acceleration due to $J_2$
for a generic orbit reads \eqi A_R^{(J_2)}=-\frac{3GMR_{\oplus}^2
J_2}{4r^4}\left[3\cos 2(\omega+f)+3\cos^2 i-3\cos^2 i\cos
2(\omega+f)-1 \right].\lb{j2i}\eqf Neglecting all terms of order
$\mathcal{O}(e^2)$, \rfr{j2i} in \rfr{longi} yields \eqi
P_l^{(J_2)}=\frac{3\pi R^2_{\oplus} J_2}{(GM
r_0)^{\frac{1}{2}}}(1-3\cos^2 i). \eqf Then, for the
counter-orbiting satellites along identical (almost) circular
orbits with inclinations $i^{(+)}\equiv i$ and
$i^{(-)}=(180^\circ-i)+I$, with $I/i\ll 1$, from $\cos^2 i\sim
1+i^2$, \eqi \Delta P_l^{(J_2)}\sim\frac{18\pi R^2_{\oplus}
J_2}{(GM r_0)^{\frac{1}{2}}}(iI).\eqf For $r_0=25498$ km and
$i=0.01^\circ$, $\Delta P_l^{(J_2)}\leq \Delta P_l^{(\rm GM)}$ if
$I\leq 0.006^\circ=1\times 10^{-4}$ rad. It is worth noting that
the current technology does allow to obtain equatorial orbits
tilted by $0.01^\circ$ to the equator, e.g., for many
geostationary satellites. Many orbital perturbations affect the
inclination with shifts $\Delta i$ which must be kept smaller than
$I$. The static part of geopotential does not induce long-term
perturbations on the inclinations. The tesseral and sectorial
solid Earth and ocean tides induces long-period perturbations on
$i$ of the order of a few tens of mas. The general relativistic
gravitoelectric de Sitter, or geodetic, precession \ct{desitter}
induces a secular variation of the inclination of 84 mas
yr$^{-1}$. The inclination is sensitive to the non-gravitational
perturbations, but the drag-free apparatus could resolve this
problem. Suffices it to say that the uncancelled non-conservative
perturbations on $i$ for the LAGEOS satellites amount to a few
tens of mas yr$^{-1}$ \ct{luc}.
It is also important to note that departures of $i$ from their
nominal values of the order of $I\sim 0.0001^\circ$ are feasible
with the current technologies for, e.g., the planned GP-B mission
\ct{axel91, pet97} and the current GRACE mission\footnote{
\texttt{http://www.csr.utexas.edu/grace/newsletter/2002/august2002.html}}.
It turns out that the requirements on the semimajor axis and the
eccentricity are far less demanding than those on the inclination
\ct{lichetal03}.

\subsubsection{The N-Body Gravitational Perturbations}

Another source of perturbations on an Earth satellite's mean
longitude is represented by the gravitational effect induced by
the major bodies of the Solar System (Sun, Moon, Jupiter, other
planets, the asteroids). Let us calculate the effects induced by
some of them.

The perturbative effect of the planet of mass $m^{'}$ on the
satellite of mass $m$ is given by\footnote{Here we adopt the
Lagrangian approach; $\mathcal{R}$ is the disturbing function and
the (approximate) equation for the rate of the mean longitude is
\eqi\dert{l}{t}=n-\frac{2}{na}\derp{\mathcal{R}}{a}.\eqf }
\ct{bocpuc98} \eqi\mathcal{R}_{\rm
planets}=Gm^{'}\left(\rp{1}{|\bf r -\bf r^{'} |}-\rp{{\bf r}\cdot
{\bf r}^{'}}{r^{' 3}}\right).\lb{bocpuc}\eqf It turns out that the
second term in \rfr{bocpuc} does not induces secular
perturbations.

After expressing the first term of \rfr{bocpuc} in terms of the
orbital elements of the satellite and the planet and averaging it
over one period of the mean longitudes $l$ and $l^{'}$ one finds
for the largest contribution which does not contain the terms of
second order in the eccentricities and the inclinations \eqi
P_l^{(\rm planets)}=-\rp{4\pi Gm^{'}}{n^3 a^{' 3}}.\lb{plan}\eqf
It is important to notice that \rfr{plan} does not depend on the
sense of motion along the orbits, so that it cancels, in
principle, in $\Delta P_l$. Instead, the difference $d$ in the
semimajor axes of the two satellites would induce an aliasing
effect \eqi\Delta P_l^{\rm (planets) }\sim\frac{18\pi
m^{'}da^{\frac{7}{2}}}{G^{\frac{1}{2}}a^{'
3}M_{\oplus}^{\frac{3}{2}}}.\lb{difplan}\eqf The nominal values of
\rfr{difplan} for the Sun and the Moon amount to 1.178$\times
10^{-4}$ s and $2.565\times 10^{-4}$ s, respectively. However, the
uncertainties in $Gm^{'}$ yield
\begin{eqnarray}
\delta[\Delta P_l^{\rm (Sun)}]_{GM_{\odot}}&=& 7\times 10^{-15}\ {\rm s},\\
\delta[\Delta P_l^{\rm (Moon)}]_{Gm_{\rm Moon}}&=& 6\times
10^{-11}\ {\rm s},
\end{eqnarray}
where we have used $\delta(GM)_{\odot}=8\times 10^{-9}$ m$^3$
s$^{-2}$ \ct{sta95} and $\delta(Gm)_{\rm Moon}=1.2\times 10^{-6}$
m$^3$ s$^{-2}$ \ct{kon}. The errors induced by the uncertainties
in $d$, $a$ and $GM_{\oplus}$ turn out to be negligible. The
corrections $\Delta P_l^{\rm (planets)}$ for Venus, Mars, Jupiter
and Ceres, whose mass amounts to approximately one third of the
total mass of asteroids $m^{\rm asteroids}=2.3\times 10^{24}$ g,
are negligible because their nominal values are $\leq 10^{-9}$ s.
This is particularly important for Jupiter, whose contribution
would amount to $\Delta P_l^{\rm (Jup)}=1\times 10^{-9} $ s,
because its sidereal orbital period, with respect to the Sun,
amount to almost 11 years.

It can be shown that the indirect effect induced on $P^{(0)}$ by
the perturbations on $a$ can be neglected. Indeed, there are no
N-body secular perturbations on $a$ and, most
importantly\footnote{After all, short-periodic terms in the
planetary longitudes would result in semi-secular signatures on
the time-scales of an Earth satellite.}, their amplitudes are
proportional to factors like $e^{|h_1|}e^{'
|h_2|}\sin^{|h_3|}i\sin^{|h_4|}i^{'}$, where $h_1,\ h_2,\ h_3$ and
$ h_4$ are integer numbers \ct{brocle} constrained  by the
conditions $|h_1|+|h_2|+|h_3|+|h_4|\leq 2$ and, for the
short--period perturbations, $h_1+h_2+h_3+h_4\neq 0$.

\subsection{The Impact of the Non-Gravitational Perturbations}

From \rfr{longi} it turns out that, in general, periodically
time-varying radial accelerations with frequencies multiple of the
orbital frequency do not affect the mean longitude over an orbital
revolution, at least at order $\mathcal{O}(e)$.

The situation is different for radial accelerations which can be
considered constant in time over an orbital revolution. Let us
consider an acceleration of non--gravitational origin whose radial
component $A_R$ is constant over the timescale of one orbital
revolution. From \rfr{longi} it turns out that \eqi P_l^{(\rm
NG)}=4\pi\left(\frac{r_0^7}{GM^3}\right)^{\frac{1}{2}} A_R,\eqf
from which it follows \eqi\Delta P_l^{(\rm NG)}=14\pi d
\left(\frac{r_0^5}{GM^3}\right)^{\frac{1}{2}} A_R.\lb{dng}\eqf The
maximum value of $A_R$ which makes $\Delta P_l^{(\rm NG)}\leq
\Delta P_l^{(\rm GM)}$ is $A_R^{\rm max}=6\times 10^{-7}$ cm
s$^{-2}$ for $d=5$ km and $r_0=25498$ km. It is not a too
stringent constraint. For LAGEOS, which is completely passive, the
largest non--gravitational acceleration, i.e. the direct solar
radiation pressure, amounts to almost $4\times 10^{-7}$ cm
s$^{-2}$; a drag-free apparatus with not too stringent
performances--well far from the $10^{-13}$ cm s$^{-2}$
Hz$^{-\frac{1}{2}}$ level of LISA--could meet in a relatively easy
way such requirement. For $\delta d\sim\delta r_0\sim 1$ cm,
$\delta(GM)=8\times 10^{11}$ cm$^3$ s$^{-2}$ and $A_R=6\times
10^{-7}$ cm s$^{-2}$ it turns out \eqia \delta[\Delta P_l^{(\rm
NG)}]_{d}&=& 14\pi\left(\frac{r_0^5}{GM^3}\right)^{\frac{1}{2}}
A_R\times (\delta d)=1\times 10^{-12}\ {\rm s},\\
\delta[\Delta P_l^{(\rm NG)}]_{r_0}&=& 35\pi d
\left(\frac{r_0}{GM}\right)^{\frac{3}{2}}
 A_R\times (\delta r_0)=5\times 10^{-16}\ {\rm s},\\
\delta[\Delta P_l^{(\rm NG)}]_{GM}&=& 21\pi d
\left(\frac{r_0}{GM}\right)^{\frac{5}{2}} A_R\times
\delta(GM)=1\times 10^{-15}\ {\rm s}. \eqfa These results show
that the direct impact of the non-gravitational accelerations on
the measurement of the gravitomagnetic time shift on $l$ could be
made negligible with a drag-free apparatus of relatively modest
performances, contrary to the indirect effects on the difference
of the Keplerian periods which, as seen before, would require a
much more effective drag--free cancellation in the frequency range
$[0,\ 10^{-5}\ {\rm Hz}]$.

%
%
%
%
%
%
\section{Conclusions}

In this paper we have examined the possibility of measuring the
gravitomagnetic clock effect on the mean longitudes of a pair of
counter-rotating satellites following almost identical circular
equatorial orbits in the gravitational field of Earth. While the
gravitomagnetic signature depends only on the Earth parameters,
the aliasing classical effects depend also on the orbital geometry
of the satellites, so that it is possible to choose it suitably in
order to reduce their impact on the measurement of the
post-Newtonian effect. In this respect, our choice of a nominal
value of 25498 km for the semimajor axis yields good results.

From the point of view of the observational accuracy, one source
of error would be the uncertainty with which the axes of the ITRS
are known. However, it turns out that the level of the
gravitomagnetic effect can be reached over a sufficiently high
number of orbital revolutions.

In regard to the systematic errors, the major limiting factors are
\begin{itemize}
  \item The unavoidable imperfect cancellation of the Keplerian periods,
  which yields a 10$^{-2}$ cm constraint in knowing the difference
  $d$ between the semimajor axes of the satellites
  \item The required accuracy in the knowledge of the inclinations
  $i$ of the satellites in presence of not exactly equatorial
  orbits. For $i=0.01^\circ$ the difference between the inclinations of the two orbital planes
  $I\equiv i^{(+)}-i^{(-)}$ should be less than 0.006$^\circ$.
\end{itemize}

A pair of drag-free ($10^{-9}$ cm s$^{-2}$ Hz$^{-\frac{1}{2}}$)
spacecraft endowed with an intersatellite tracking apparatus might
allow to meet the stringent requirements of such a mission whose
realization seems to be very difficult although, perhaps, not
completely impossible with the present-day or forthcoming space
technologies.

\section*{Acknowledgments}
L.I. is grateful to L. Guerriero for his support while at Bari.
L.I. and H. L. gratefully thanks the 2nd referee for his/her
useful and important efforts for improving the paper.

\end{document}